\documentclass[runningheads]{llncs}
\usepackage[T1]{fontenc}

\usepackage{pgfplotstable}

\usepackage[colorinlistoftodos]{todonotes}
\usepackage[cmex10]{amsmath}
\usepackage{xspace}
\usepackage{xcolor}
\usepackage{cite}
\usepackage[hyphens]{url}

\usepackage{graphicx,subcaption}
\usepackage{balance} 
\usepackage{listings} 
\usepackage{pifont}% http://ctan.org/pkg/pifont\usepackage{lipsum}
\usepackage{lipsum} 
\usepackage{multirow} 
\usepackage{booktabs}
\usepackage[linesnumbered,figure]{algorithm2e}

\usepackage{pgfplots}
\usepackage{tikz}
\usepackage{tablefootnote}

\usepackage{ifpdf}

\usepackage{textcomp}
\usepackage{spverbatim}
\usepackage{balance}
\usepackage{amssymb}% http://ctan.org/pkg/amssymb
\usepackage{pifont}% http://ctan.org/pkg/pifont\usepackage{lipsum}
\usepackage{fancybox}
\usepackage{lipsum}
\usepackage{ifthen}
\usepackage{listings}
\usepackage{multirow}
\usepackage{colortbl}
\usepackage{tabularx}
\usepackage{booktabs}

\definecolor{tablegray}{RGB}{200,200,200}
\definecolor{myGreen}{HTML}{006000}

\definecolor{shadecolor}{rgb}{1,0,0} % this is blue
\newboolean{showcomments}
\setboolean{showcomments}{false}   % deactivate
\setboolean{showcomments}{true}    % activate

\definecolor{gray}{gray}{0.8}
\newcommand{\boxfix}[2][]{%
  \ifthenelse%
    {\boolean{showcomments}}%
    {\colorbox{gray}{\parbox{0.95\linewidth}{\emph{\normalsize\sffamily{\bfseries\/Comment\ifthenelse{\equal{#1}{}}{:}{~#1:}}~#2}\/}}}%
    {}%
}

\usepackage{listings}
\newcommand\blfootnote[1]{%
  \begingroup
  \renewcommand\thefootnote{}\footnote{#1}%
  \addtocounter{footnote}{-1}%
  \endgroup
}

\begin{document}

\title{Representing LLVM-IR in a Code Property Graph}

\author{
  Alexander K\"uchler\orcidID{0000-0001-6633-460X}
  \and
  Christian Banse\orcidID{0000-0002-4874-0273}
}

%\authorrunning{A. K\"uchler \and C. Banse}

\institute{
  Fraunhofer AISEC, Germany\\
  \email{\{alexander.kuechler,christian.banse\}@aisec.fraunhofer.de}
}

\maketitle

\begin{abstract}
  In the past years, a number of static application security testing tools have
  been proposed which make use of so-called code property graphs, a graph model
  which keeps rich information about the source code while enabling its user to
  write language-agnostic analyses. However, they suffer from several
  shortcomings. They work mostly on source code and exclude the analysis of
  third-party dependencies if they are only available as compiled binaries.
  Furthermore, they are limited in their analysis to whether an individual
  programming language is supported or not. While often support for
  well-established languages such as C/C++ or Java is included, languages that
  are still heavily evolving, such as Rust, are not considered because of the
  constant changes in the language design.
  To overcome these limitations, we extend an open source implementation of a
  code property graph to support LLVM-IR which can be used as output by many
  compilers and binary lifters. In this paper, we discuss how we address
  challenges that arise when mapping concepts of an intermediate representation
  to a CPG. %, which is mostly focused on higher-level programming languages. 
  At the same time, we optimize the resulting graph to be minimal and close to
  the representation of equivalent source code. Our evaluation indicates that
  existing analyses can be reused without modifications and that the performance
  requirements are comparable to operating on source code. This makes the
  approach suitable for an analysis of large-scale projects.
  \blfootnote{This preprint has not undergone peer review (when applicable) or
  any post-submission improvements or corrections. The Version of Record of this
  contribution is published in \textit{25th International Conference on
  Information Security (ISC)}, and is available online at
  \url{https://doi.org/10.1007/978-3-031-22390-7_21]}.}
\end{abstract}

\section{Introduction}
Despite huge efforts of researchers and industry put into identifying vulnerable
software, %and providing program analysis tools,
many software systems still suffer from various security weaknesses.
%However, in
%many scenarios, secure and bug-free software is highly critical. Some weaknesses
%are hard to identify because many industry-standard analysis tools cannot detect
%them correctly, others range from the ever-growing complexity of software and
%its interactions with other programs, yet other weaknesses might not be
%detectable by a tool if the tool fails to provide the right information.
The concept of a code property graph (CPG) \cite{yamaguchi2014modeling} has been
introduced to simplify the identification of vulnerabilities and bugs in the
source code of programs. A CPG is a super-graph covering properties of an
abstract syntax tree (AST), a control flow graph and data flow graphs, among
others, thus containing all information relevant for a security analysis. The
CPG enables its user to identify vulnerabilities or bugs by performing reusable
graph queries. This perk led to a widespread adaption of the technique with
several implementations \cite{xiaomeng2018cpgva,yamaguchi2014modeling,Graft,Plume,Joern,CPG,banse2021cloud,weiss2022languageindependent}.
%Furthermore, those graph queries are ideally easy to write, thus allowing a
%targeted identification of application-specific bugs.
Even if the graphs mimic the source code with a minimal loss of information, the
graph provides an abstraction of the actual code. This abstraction is
suitable to support a language-agnostic analysis of software.
%which is why many tools can handle multiple programming languages.
Unfortunately, the implementations are limited with respect to the %number of
supported programming languages since each language requires a separate
translation.
%In the past years, the concept has gained importance and has been implemented by
%several analysis tools \cite{xiaomeng2018cpgva,yamaguchi2014modeling,Graft,Plume,
%Joern,CPG,banse2021cloud,weiss2022languageindependent}. Each of the tools has
%different applications and, depending on their use-case, can include additional
%data. The resulting graphs mimic the source code of the program with the least
%possible loss of information while still providing an abstraction of the actual
%code. Furthermore, the CPG representation turned out to be a suitable approach
%to enable a language-agnostic analysis of software artifacts which led to tools
%which can handle multiple programming languages. Unfortunately, those approaches
%are limited with respect to their support for different programming languages
%and need a separate translation step for each supported language.

As compilers suffer from a similar problem, the use of intermediate
representations (IR) has become popular. The IR abstracts from the programming
language but, in many cases, still contains a significant amount of high-level
information such as the types of variables which is lost in the compiled binary
and can barely be recovered \cite{mantovani2022the}.
%This is interesting for the analysis
%because such information is lost in the compiled binary and can barely be
%recovered \cite{mantovani2022the}.
The lack of such information can worsen the analysis results. A very
popular IR is LLVM-IR \cite{llvmir} which is part of the LLVM
project. Numerous compiler frontends exist to translate programming languages to
LLVM-IR. E.g., clang \cite{clang} translates the
languages C, C++ and Objective-C to LLVM-IR and has been extended by Apple to
support Swift \cite{swiftClang}. Other frontends exist to support a wide range
of programming languages (e.g., Rust).

While LLVM-IR was designed for compilers, it is also frequently used by
binary lifters. E.g., RetDec \cite{RetDec}, McSema
\cite{mcsema}, llvm-mctoll \cite{mctoll} and reopt \cite{reopt} can
lift a binary to LLVM-IR. Since some lifters
support multiple architectures and types of binary files, this avoids to
implement the translation for different flavors of assembly code or application
binary interfaces.
%which can differ significantly among different target platforms.
Recent research \cite{liu2022sp} showed that binary lifting is meanwhile a
stable technique and its produced output is suitable for a security analysis of
the program. Another use-case for analyzing LLVM-IR is to consider the effects
of compiler optimizations during the analysis. As an example, recent research
showed that side channel vulnerabilities introduced by the compiler are still a
major concern of developers of cryptographic libraries \cite{jancar2022sp} and
that the source code and the final binary files can differ significantly
\cite{balakrishnan2010wysinwyx}. Since the LLVM-IR can be emitted after the
optimizations, it can already contain the vulnerabilities or bugs which stem
from the compiler and is therefore an interesting analysis target.

In this paper, we present an approach how to overcome shortcomings of existing
CPG tools by enabling the analysis of LLVM-IR in a code property graph. This
bridges the gap between the analysis of source code written in higher-level
programming languages and the analysis of programs (or dependencies) that may
only exist in binary form. While supporting LLVM-IR in a CPG seems to be
straightforward, several challenges arise from the static single assignment
(SSA) form, the exception handling routine, instructions which do not exist in
high-level programming languages and significantly different syntactic
representations of some concepts in LLVM-IR and other languages. Contrary to
prior work, we do not require to run any LLVM passes beforehand, which helps us
to keep the graph smaller. At the same time, we aim to retrieve as much
high-level information as possible and map the code to high-level concepts
whenever possible. Rather than handling LLVM-IR-specific instructions, e.g.,
\textit{cmpxchg}\footnote{The \textit{cmpxchg} instruction compares a given
  argument against a value stored in a memory address. If they are equal, a new
  value, specified in a second argument is stored in memory. This is similar to
  \texttt{if(*addr == arg0) {*addr = arg1;}} in C/C++.}, only as a generic
function call, we translate the concepts into existing CPG node types that
represent the behavior of a higher-level programming language. This allows us to
re-use existing concepts in queries, such as if-statements or pointer
referencing. Overall, integrating LLVM-IR into a CPG allows to support more
programming languages, to analyze binary files and to validate that a compiler
did not introduce new bugs. It enables that existing analyses queries for source
code can be applied to the LLVM-IR without any modifications. In summary, our
contributions are as follows:
\begin{itemize}
  \item We are the first to present a mapping of all LLVM-IR instructions to
        existing CPG nodes with full compatibility to the existing structure.
        This ensures that existing analyses are fully compatible with the
        representation.
  \item We show how we can keep the size of the CPG minimal.
  \item We are the first to include LLVM-IR's exception handling routines in a CPG.
  \item We extended the open source project \textit{cpg}
        \cite{weiss2022languageindependent,CPG} to support our concepts.
\end{itemize}

%The rest of this paper is structured as follows. Section \ref{sec:background}
%provides the required background. Section \ref{sec:related_work} summarizes
%related work concerning CPG-based analyses. Section \ref{sec:mapping} gives
%details on how we map the LLVM-IR to the constructs in the CPG, while
%\ref{sec:pass} shows how we compress the resulting graph to mimic the properties
%of programs present in source code more closely compared to the ones given in
%LLVM-IR. Section \ref{sec:eval} presents a case study which shows how the
%resulting graph can be used to detect cryptographic misuses and that the
%LLVM-based representation is suitable to reuse the same query used for the
%source code of the program. The experimental evaluation further indicates that our resulting graph does
%not lead to an explosion of the nodes and edges. Section \ref{sec:discussion}
%discusses our results and outlines future research directions and application
%areas. Finally, Section \ref{sec:conclusion} concludes the paper.

\section{Background}
\label{sec:background}
%In this section, we summarize the background on the \textit{cpg} project and
%LLVM-IR.

\subsection{The Code Property Graph}
The \textit{cpg} project \cite{weiss2022languageindependent,CPG} enables a
graph-based representation of
source code of different programming languages. To date, the focus lies in
Java and C/C++ but experimental support for Python, Go and TypeScript is also
available. The goal of the project is to provide a language-agnostic
representation of the source code. This enables a security expert to identify
vulnerabilities or bugs.
Furthermore, the \textit{cpg} library comprises a way to store the graph in
neo4j\footnote{\url{https://neo4j.com/}}, and makes the graph
accessible via a command line interface.
For some cases, the library can also evaluate the value
which can be held by a node. All this allows a security expert to write custom
queries either to the graph database or the in-memory representation of the CPG.
The \textit{cpg} library is designed in a way to allow reusing these
queries among all supported programming languages.
To fulfill this goal, the \textit{cpg} library implements a thorough class hierarchy which
accounts for various types of statements and expressions. The CPG encodes
information such as the class hierarchy of the code under analysis, the control
flow graph, and the call graph in a single graph. The current design
mainly targets object-oriented programming languages. To deal with a possible
lack of some code fragments or errors in the code, the library is resilient to
incomplete, non-compilable and to a certain extent even incorrect code.

%Furthermore, the
%\textit{cpg} makes several assumptions on valid code, such as the presence of
%exactly one declaration of every variable.

\subsection{The LLVM Intermediate Representation}
%In this section, we summarize the required background on the LLVM-IR.

\noindent\textbf{The Instructions.}~
The LLVM-IR is used as IR of the LLVM project. Its main
purpose lies in providing an abstraction of code %during compilation
to ease the optimization and analysis of the program in a language-
and architecture-independent way. The LLVM-IR holds values in global variables
(prefixed with \texttt{@}) and local variables (prefixed with \texttt{\%})
both of which can be named or unnamed. The LLVM-IR follows the static single
assignment (SSA) form. Hence, every variable can be written to exactly
once. This limitation does not affect global variables as they are represented
as memory locations and are accessed via store or load operations.

Overall, %the instruction set is kept simple;
the LLVM-IR differentiates between 65
instructions. Of these, 13 are arithmetic operations, 6 are bitwise
operations, and 13 instructions cast types. The remaining instructions call
functions, handle exceptions, load from or store to memory,
manipulate aggregated types or jump to other program locations. %This limited
%set of instructions is enhanced with numerous metadata which can be used
%for optimizations, to annotate the calling convention, desired properties of
%functions and parameters, among others.
The instructions can be enhanced with metadata to note the calling
convention, optimizations or desired properties of functions and parameters,
among others.

Besides the basic instructions, LLVM-IR contains numerous so-called
``intrinsics''. Those are functions which model certain standard library
functionality, or model frequent actions which have to be represented
differently on different architectures. Some intrinsics repeat or refine basic
instructions, others insert functionality such as the automated memory
management in Objective-C.

The LLVM-IR supports a simple type system and differentiates between a set of
primitive types %(integers and floating-point values)
and aggregated types such as
structs, arrays and vectors. Additionally, LLVM-IR has a type for
labels (i.e., jump targets), metadata and a so-called token which is used by
certain instructions to transport information. Overall, the type system
resembles C rather than object-oriented programming languages. In fact,
object-oriented concepts are handled by the respective language
frontend in LLVM. The frontend translates the object-oriented properties to
concepts such as VTables for overriding methods, and method
name mangling to support overloaded functions. In the case of
Objective-C, it uses the dynamic dispatching strategy. Other languages make use
of similar concepts.

\noindent\textbf{Accessing LLVM-IR.}~
The LLVM project offers a C++ and a C API to parse LLVM-IR and
LLVm bitcode files.
% Bitcode files hold the information in a compressed format and not a textual representation.
As the CPG project is mainly implemented in
Java, access to the API has to be provided via the Java Native Interface (JNI).
We use the open source project
javacpp-presets\footnote{\url{https://github.com/bytedeco/javacpp-presets/tree/master/llvm/src/gen/java/org/bytedeco/llvm}}
which provides access to the C API via JNI. Unfortunately, the C API has a flat
type hierarchy in its functions to access the LLVM-IR's AST, thus making the parsing
of instructions and the extraction of their elements more error-prone if not
parsed correctly\footnote{Typically, an incorrect API call leads to a segfault.}
%because the C code would try to access invalid data}.
However, as our evaluation
in Section \ref{sec:eval} shows, our implementation works in a stable way.

\section{Related Work}
\label{sec:related_work}
\noindent\textbf{Code Property Graphs.}~
Researchers and industry proposed multiple use cases and implementations of CPGs
and analysis tools \cite{xiaomeng2018cpgva,click1995a,yamaguchi2014modeling,Graft,Plume,Joern,CPG,banse2021cloud,weiss2022languageindependent,schuette2019lios}. All of these tools differ in
their support for programming languages.

Closest to our work is the tool \texttt{llvm2cpg} \cite{llvm2cpg} which uses
Joern \cite{Joern} as graph representation.
The respective CPG represents most instructions as function calls and
does not try to infer any of the high-level information. Furthermore, it uses
the \textit{reg2mem} LLVM pass to address the $\varphi$ instruction of LLVM-IR,
which significantly increases the code base. This results in additional
instructions present in the graph and thus slows down the analysis
and makes it more error-prone.

liOS \cite{schuette2019lios} constructs
a CPG holding assembly instructions and the function
bodies lifted to LLVM-IR to analyze iOS apps. The graph model cannot be used to
represent source code. Furthermore, liOS does not
specifically address LLVM-IR instructions since the analyses mainly operate on
assembly code.

Plume \cite{Plume} and Graft \cite{Graft,keirsgieter2020graft} only support
Java bytecode, a different low-level language. Plume
builds the graph incrementally to analyze data flows and has been merged
into Joern in a revised version. Graft follows a similar goal.
Other tools \cite{click1995a,yamaguchi2014modeling,Joern,CPG,weiss2022languageindependent} analyze source code and differ in their level of
abstractions and supported languages. Some tools extend CPGs for specific use
cases, e.g., analyzing cloud apps \cite{banse2021cloud} or finding
vulnerabilities with deep learning \cite{xiaomeng2018cpgva}.

\noindent\textbf{Graph-based Security Analysis.}~
Various other works investigated in the usage of other graph-based
representations of the source code to identify bugs or vulnerabilities
\cite{urma2015source,yamaguchi2012generalized,yamaguchi2015automatic} or
similar code fragments\cite{gascon2013structural,baxter1998clone},
traverse the graph \cite{Rodriguez2015} or improve the analysis
\cite{lam2005context}. These works aim to provide a rich basis for analyzing the
graphs. Many of the proposed techniques operate on other graph structures (e.g.
the AST). However, the CPG combines a multitude of information and includes the
respective relations, thus making the required information available for the
analysis. Hence, these approaches can still be applied to the CPG.% even if they
%had not been tailored to it specifically.

\noindent\textbf{Static Analysis of Multiple Programming Languages.}~
Other works target the analysis of multiple programming languages
%via different techniques
\cite{caracciolo2014pangea,flores2015cross,flores2011towards,angerer2014variability,mushtaq2017multilingual,mayer2012cross}.
Some of the frameworks rely on language-agnostic ASTs
\cite{schiewe2022advancing,zugner2021language} or aim to provide a
common pattern for the AST of multiple languages \cite{rakic2013language,strein2006cross}. However, ASTs are cannot be used to find all kinds of bugs
as they do not contain the required information \cite{yamaguchi2014modeling}.
Teixeira et al. \cite{teixeira2021multi} even propose to translate source code
to a custom language.

Furthermore, various intermediate representations (IRs) have been proposed
either for compilers (e.g., LLVM \cite{lattner2004llvm}, GIMPLE \cite{gimple},
HIR \cite{hir} or CIL \cite{ecmaecma}), or specifically targeting code analysis
(e.g. VEX IR \cite{nethercote2007valgrind,vex}, jimple \cite{vallee1998jimple},
BIL \cite{brumley2011bap}, REIL \cite{dullien2009reil}, ESIL \cite{esil}, DBA
\cite{bardin2011bincoa,david2016binsec} or RASCAL \cite{klint2009rascal}).
Since the IRs are often tailored to a specific use case or language, they
differ in the information available in the instructions and their abstractions.
Many of the IRs are integrated in analysis toolchains whose analyses are often
specific to a use case and cannot easily be ported to other tools. Therefore,
integrating such IRs in an abstract analysis platform like the CPG can enable
further generalized security analysis.
%The IRs differ in their levels of abstraction, the information available in the
%instructions and are often tailored to a specific use case (e.g. binary analysis
%for BIL, ESIL or DBA) or language (e.g. Java bytecode for jimple).
%Furthermore,
%while many of the IRs have been developed as part of an analysis toolchain, the
%respective analyses are typically still highly specific to a use case.
%Therefore, integrating such IRs in an even more abstract analysis platform such
%as the CPG can enable further generalized security analysis which are not
%included in the original toolchain.

Numerous tools \cite{fbinfer,sonarqube,checkmarx,appscreener,codacy,codeql,avgustinov2016ql,de2007keynote,codechecker,coverity,deepsource,lgtm,wala} can analyze multiple programming
languages. %and are often integrated in the CI/CD pipeline of the development.
However, they can often barely share the analyses between the %programming
languages.
%However, typically, they cannot or barely use the analysis of one language for
%another one, and often are commercial software.
%Furthermore, most of these tools suffer from limited analysis  capabilities.
The CPG representation allows reusing analyses across languages.
%Furthermore, the tool is open-source.

\section{Mapping LLVM-IR to CPG nodes}
\label{sec:mapping}
We aim to include LLVM-IR in the CPG while reusing only the existing node types
and representing LLVM-specific constructs similar to their equivalents in %programming
languages which are already supported by the CPG.
We also want to keep the number of nodes minimal.
%The main challenge of this task is to map all LLVM-IR instructions to matching CPG nodes.
In this section, %we show how we perform this mapping from the low-level
%instructions in the LLVM-IR to the high-level concepts in the CPG. Specifically,
%we explain how we map the arithmetic and logical instructions of
%LLVM-IR, how we can handle aggregate types in a way to reuse existing node
%structures, how we represent the $\varphi$ instruction with a minimal increase of
%nodes and how we aim to handle LLVM-IR's exception handling instructions.
we present how we represent 1) arithmetic and logical instructions, 2) access
to aggregate types, 3) the $\varphi$ instruction with a minimal increase of
nodes, and 4) LLVM-IR's exception handling routine.

\subsection{Basic Instructions}
Many instructions are known from other programming languages. We can coarsely
differentiate between arithmetic and logical operations, operations which
enforce specific interpretations of types, and operations which are composed of
numerous steps but are often performed atomically on the CPU. In this section,
we explain how we include those respective instructions in the CPG.

Almost all programming languages have a common subset of instructions or
operations. This includes arithmetic, bitwise and logic operations, or
comparisons which we map to their representation in high-level
languages (\texttt{+}, \texttt{-}, \texttt{*}, \texttt{/}, \texttt{\%},
\texttt{$\hat{ }$} , \texttt{\&}, \texttt{|}, \texttt{<<}, \texttt{>>},
\texttt{<}, \texttt{<=}, etc.).
Other instructions like jumps, calls, return instructions are modeled with
their representation in C code. For if- and switch/case-statements,
the branches or cases contain a simple goto statement. Later, a CPG pass
removes such indirections whenever possible to reduce the size of the graph.

For some instructions, LLVM-IR can enforce a specific interpretation of the
types of the arguments. E.g., the instructions \texttt{udiv},
\texttt{sdiv} and \texttt{fdiv} represent a division and are mapped to the
binary operator \texttt{/}. However, they interpret the values as unsigned
(\texttt{udiv}), signed (\texttt{sdiv}) or as floating point value
(\texttt{fdiv}). In the CPG, we add typecasts to the arguments to enforce the
correct interpretation.
%LLVM-IR differentiates between different data types. We model those
%differences by adding type casts to the respective operands. As an example, the
%instructions \texttt{udiv}, \texttt{sdiv} and \texttt{fdiv} all represent a
%division and thus are mapped to the binary operator \texttt{/}. However, the
%operator \texttt{udiv} treats the values as unsigned, which makes us add a type
%cast to both operands. \texttt{sdiv} treats the value as signed and
%\texttt{fdiv} as floating point value. As these are the default interpretations
%(for integer or floating point value), we do not have to add typecasts.

%Apart from operators holding type information or how to interpret a certain value,
In addition,
some comparators of floating point values
check if a number is ordered or not (i.e., if it is \texttt{NAN}).
We split these comparisons into a check if
the number is ordered and then the actual comparison. E.g., the
comparators \texttt{ult} and \texttt{olt} %in an \texttt{fcmp} instruction
compare two floating point values and are mapped to the \texttt{<} operator.
However, the \texttt{ult} comparison checks if a value \texttt{a} is unordered
or less than value \texttt{b} and thus is modeled
as the statement \texttt{std::isunordered(a)||a<b}.
Similarly, we model the \texttt{olt} comparison with
\texttt{!std::isunordered(a)\&\&!std::isunordered(b)\&\&a<b}.

Some of LLVM's instructions like \texttt{atomicrmw} and \texttt{cmpxchg} are
known from assembly code rather than high-level languages and perform multiple
operations atomically.
The \texttt{cmpxchg} instruction loads a value from memory and
replaces it with an operand if the value equals another operand. In the CPG,
we model this by a block of statements holding the comparison, an if statement and
the assignment in the then-branch.
We annotate the block to keep the information that all this is
performed atomically. Similarly, we model \texttt{atomicrmw} as a block of
statements performing a load, an optional comparison and if-statement and an
assignment to a variable.
By modeling these instructions with a representation similar to
source code, we simplify subsequent analyses. In contrast, prior work
\cite{llvm2cpg} models these instructions as a call to custom functions.

\subsection{Handling Aggregate Types}
High-level languages provide syntactic means to access elements of complex types
like arrays, structs or objects. In LLVM-IR, arrays and structs are still
present and their values can be accessed by special instructions. For arrays
which are represented as a vector, the instructions \texttt{extractelement} and
\texttt{insertelement} provide access to the elements.

\begin{figure}[t]
  \centering
  \includegraphics[width=0.78\columnwidth]{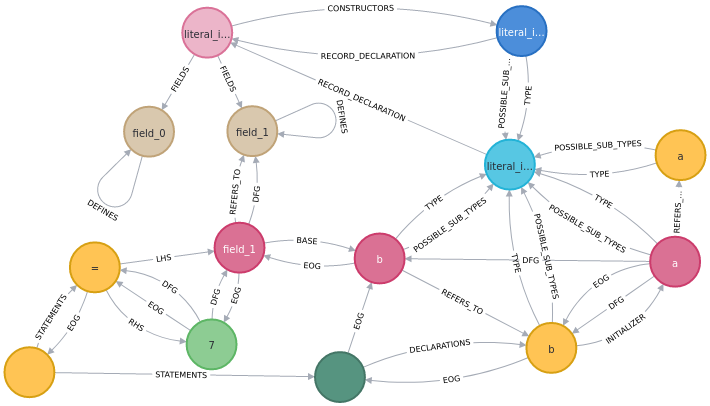}
  \caption{The graph representing the insertvalue instruction. We can see the
    literal struct which is generated as well as the access to the field.}
  \label{fig:insertvalue}
\end{figure}

Both instructions are represented as an \texttt{ArraySubscriptionExpression} in
the CPG, one being the left-hand side of the assignment and one the right-hand
side. Note that \texttt{insertelement} returns the modified vector and does not
modify the existing one. For all other aggregated types, the instructions
\texttt{getelementptr}, \texttt{extractvalue}, and \texttt{insertvalue} model
the access to the element either by the index inside an array or by the position
of a field inside a structure. The code \texttt{\%b = insertvalue {i32, i8} \%a,
  i8 7, 1} shows how the second element of the variable \texttt{a} is set to
\texttt{7}.
%The example shows an interesting property. In LLVM-IR the layout
%of a struct can be specified in the instruction. For these so-called literal
%structures, we generate a custom type which is identified by the types of the
%fields. They are unique based on their fields, which means that two literal
%structs containing the same list of fields are regarded as the same type.
%In our example, we name the struct \texttt{literal\_i32\_i8} and can replace all
%occurrences of such a structure. We model the respective struct with two fields
%(\texttt{field\_0} of type \texttt{i32} and \texttt{field\_1} of type
%\texttt{i8}).
We model the instruction as a copy of \texttt{a} to the variable
\texttt{b} and an assignment of the value \texttt{7} to the accessed
\texttt{field\_1}.
Figure \ref{fig:insertvalue} shows the resulting graph with the initialization
of \texttt{b} on the bottom right, and the access to the field on the left.
%The variable \texttt{b} (bottom right) is first
%initialized with the value of \texttt{a}. Then, on the left side of the graph,
%the value of \texttt{field\_1} is set to 7.

\begin{figure}[htb!]
  \centering
  \begin{subfigure}{\textwidth}
    \includegraphics[width=\columnwidth]{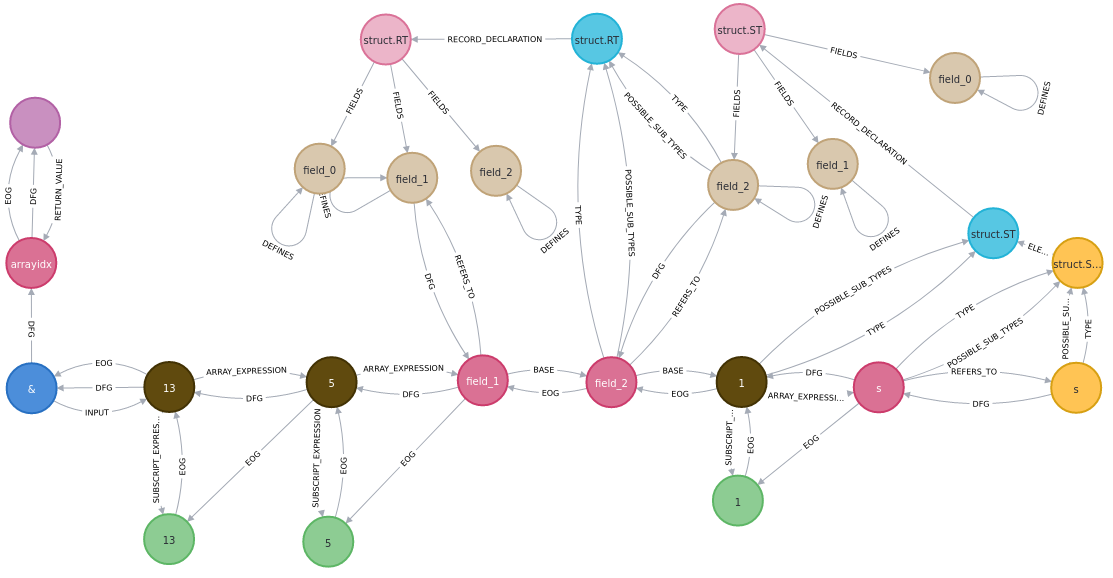}
    \caption{The graph representing the getelementptr instruction.}
    \label{fig:getelementptr}
  \end{subfigure}

  \begin{subfigure}{\textwidth}
    \includegraphics[width=\columnwidth]{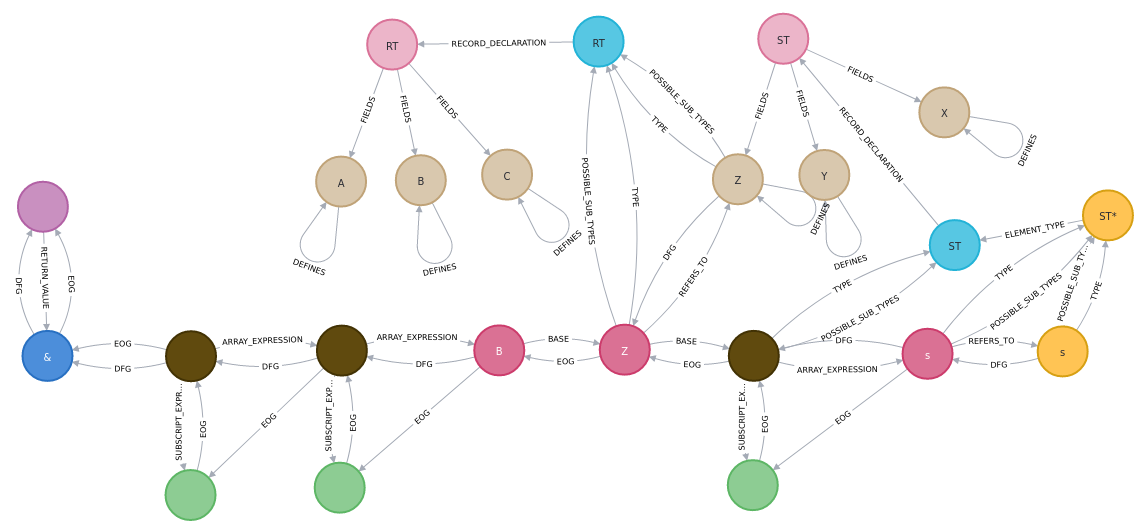}
    \caption{The graph representing the C code.}
    \label{fig:cGetelementptr}
  \end{subfigure}
  \caption{Comparison of the LLVM code using getelementptr and the respective
    C code. The graph contains structs (light pink) and their fields (light
    brown), the access to fields (dark pink), access to elements in arrays
    (brown) and the return instruction (purple). The green nodes are constant
    values, the yellow node is the method's argument. The structure of both
    graphs is nearly identical.}
\end{figure}

The example uses an interesting concept of LLVM-IR: a
so-called literal structure, a struct whose layout is
defined in the instruction. For such structs, we generate a type which is
identified by the types of its fields. Hence, all literal structs
with the same list of fields are regarded as the same type.
In our example, the struct is named \texttt{literal\_i32\_i8} and has the
fields \texttt{field\_0} of type \texttt{i32} and \texttt{field\_1} of type
\texttt{i8}.
The top left of Figure \ref{fig:insertvalue} shows the declaration of the
type.
%\noteAlex{The ``copy'' is currently only a normal decl +
%assignment. Maybe we should make a memcpy or so?}
%\begin{figure}[tb] %Float somewhere!
%  \begin{minipage}{0.9\linewidth}
%    \begin{lstlisting}[caption={Changing a value of a literal struct},label={lst:insertvalue}]
%%b = insertvalue {i32, i8} %a, i8 7, 1
%\end{lstlisting}
%  \end{minipage}
%\end{figure}
While the instructions \texttt{insertvalue} or \texttt{extractvalue}
read or write values from memory, it is sometimes desirable to retrieve a pointer
to an element of a structure. For this case, the instruction
\texttt{getelementptr} computes a memory address without accessing
memory. Listing \ref{lst:getelementptr} illustrates the usage of this
instruction on a named struct. Listing \ref{lst:cGetelementptr}, in turn, shows
the same code written as C.
% and shows how this instruction can be used to
%model access to identify the location of the respective element inside the array.
Figure \ref{fig:getelementptr} shows the definition of the named struct and the
connections between the fields for the graph retrieved from  LLVM-IR.
The result is remarkably similar to the graph in Figure \ref{fig:cGetelementptr}
which represents the C code.
This similarity lets us reuse existing
analyses for the graphs retrieved from LLVM-IR and shows that the graphs
are structurally identical.
In fact, the relations between variables and fields could be better resolved
which can lead to improved analysis results.
%In fact, some edges between variables and the fields
%are present only in the graph for LLVM-IR which indicates a better resolution of
%those relations which can lead to improved analysis results.

%\begin{figure}[tb] %Float somewhere!
\begin{minipage}{\linewidth}
  \begin{lstlisting}[caption={The instruction getelementptr for a named struct},label={lst:getelementptr}]
%RT = type { i8, [10 x [20 x i32]], i8 }
%ST = type { i32, double, %RT }
define i32* @foo(%ST* %s) {
  %arrayidx = getelementptr inbounds %ST, %ST* %s,
    i64 1, i32 2, i32 1, i64 5, i64 13
  ret i32* %arrayidx
}
\end{lstlisting}
\end{minipage}
%\end{figure}

%\begin{figure}[tb] %Float somewhere!
\begin{minipage}{\linewidth}
  \begin{lstlisting}[caption={The C code for the example in Listing \ref{lst:getelementptr}},label={lst:cGetelementptr}]
struct RT { char A; int B[10][20]; char C; };
struct ST { int X; double Y; struct RT Z; };
int *foo(struct ST *s) {return &s[1].Z.B[5][13];}
\end{lstlisting}
\end{minipage}
%\end{figure}

\subsection{The $\varphi$-Instruction}
The SSA form enforces that each variable is assigned exactly once in LLVM-IR.
However, in some cases, it is required to assign a value multiple times.
A frequent example is a loop counter which is set before executing the
loop and is updated on each iteration. %Another example is a variable which
%gets assigned different values in the then- and the else-branch of an
%if-statement.
%
To allow such behavior without duplicating code and without storing the values
in memory, the $\varphi$-instruction is used.
%LLVM-IR has a special instruction, the $\varphi$-instruction.
It assigns the target variable one of the inputs based on the
previously executed basic block (BB). %This allows for assigning the variable
%depending on the control flow with only a single assignment.

As most programming languages do not have such an instruction, there is no
fitting node to represent this in the CPG.
%Also, subsequent analyses do not handle such a pattern.
To address this issue, prior work \cite{llvm2cpg} relied on the LLVM
reg2mem pass\footnote{\url{https://llvm.org/doxygen/Reg2Mem_8cpp_source.html}} %\cite{reg2mem}
which translates the instruction to multiple load
and store operations. However, this pass also transforms the access to other
variables and thus significantly increases the size of the resulting CPG.
As this reduces the scalability of subsequent analyses, we avoid this
LLVM pass. %As the CPG does not require the SSA-form, 
We collect all
$\varphi$-instructions during the translation. Finally, we parse the
instructions to identify the predecessor BBs and
add an assignment to the target variable at the end of the BB. To
keep the CPG clean, we further insert a declaration of the variable at the %very
beginning of the function containing the $\varphi$-instruction and all
BBs%. This is required as the CPG and its passes expect exactly one
%declaration for each variable
\footnote{For all other variables, the statement of the assignment performs the
  declaration.}. This, however, breaks the SSA form. The snippet in Listing
\ref{lst:phi} contains the $\varphi$-instruction while Listing \ref{lst:phiAfter} shows the
function's model in the CPG.

\begin{figure}[tb]
  \noindent\begin{minipage}{.45\textwidth}
    \begin{lstlisting}[caption={Code snippet using the $\varphi$-instruction},label={lst:phi}]
define i32 @main(i32 %x) {
  %cond = icmp eq i32 %x, 10
  br i1 %cond, label %BB1,
    label %BB2
BB1:
  %a = mul i32 %x, 32768
  br label %BB3
BB2:
  %b = add i32 %x, 7
  br label %BB3
BB3:
  %y = phi i32 [ %a, %BB1 ],
    [ %b, %BB2 ]
  ret i32 %y
}
\end{lstlisting}
  \end{minipage}
  \hfill
  \begin{minipage}{0.45\textwidth}
    \begin{lstlisting}[caption={Snippet using the $\varphi$-instruction as modeled in the CPG},label={lst:phiAfter}]
define i32 @main(i32 %x) {
  ; VariableDeclaration of %y
  %cond = icmp eq i32 %x, 10
  br i1 %cond, label %BB1,
    label %BB2
BB1:
  %a = mul i32 %x, 32768
  %y = %a
  br label %BB3
BB2:
  %b = add i32 %x, 7
  %y = %b
  br label %BB3
BB3:
  ret i32 %y
}
\end{lstlisting}
  \end{minipage}
\end{figure}

\subsection{Exception handling}
LLVM-IR offers a rich system for exception handling. The CPG
represents exception handling routines with try-catch
statements. To make the LLVM-IR fit into this pattern, we need to identify which
instructions form a try-block and which ones a catch-block.
Concerning the try-block, we represent the \texttt{invoke} instruction
as a try-block surrounding a function call and a goto-statement.

For the catch-blocks, however, such a straightforward model is not possible.
In LLVM, the \texttt{catchswitch} instruction selects a matching
\texttt{catchpad} based on the signature of the catchpad-instruction of a basic
block. The catchpad contains the code of the catch-block and is ended by a
\texttt{catchret} instruction. However, the matching and signature cannot easily
be transferred to a high-level name. Therefore, we model this construct as a
catch-block which catches all exceptions and contains if-statements representing
the signature matching. If none of them matches, the exception is thrown again.
The remaining constructs such as the \texttt{cleanuppad} and its
\texttt{cleanupret} instruction are not modeled specifically.

Another way to
mark a catch-block is the \texttt{landingpad}-instruction which, again, filters
for the right object to catch. Once more, the matching is specific to the
programming language and thus, modelling this is left to future work. If we
cannot translate the instructions to concepts supported by the CPG, we model
them as special functions similar to the LLVM intrinsics.

\section{LLVM-Specific CPG Passes}
\label{sec:pass}
During the translation of the LLVM instructions to CPG nodes, the frontend
generates various instructions which later turn out to be unnecessary and thus
can be removed. This clean-up phase takes place in a pass over the CPG nodes.

First, none of the conditional jumps and switch/case-statements incorporates a
meaningful body of statements. Instead, they are modeled as goto
statements to another basic block. The pass identifies all basic blocks which
have only a single predecessor and replaces
the respective goto-statement with the basic block. Note that we do not perform
this transformation if multiple predecessors exist because it would
unnecessarily increase the number of nodes in the graph.

Second, the pass removes the instructions which serve as intermediate
steps during the generation of catch-blocks and propagates the caught exception
to the final throw statement if none of the
catchpad instructions matches.

As we explicitly aim to handle lifted or decompiled code, a second pass can
remove method stubs, i.e., methods whose only purpose is to call a library
method. The main purpose of this pass is to simplify subsequent analyses.

\section{Experimental Evaluation}
\label{sec:eval}
To reuse the same analyses for the graphs constructed from source code as well
as the ones containing LLVM-IR, we carefully designed the translation in a way
to mimic the concepts used in source code as closely as possible. In this
section, we first show a case study which advocates that we can reuse queries that
aim to identify security concerns in source code to query LLVM-IR. Second, we
test the implementation against the Rust standard library to
show the applicability of the approach to large-scale projects.
All measurements were performed on a Ubuntu 20.04 running on an Intel i5-6200U
CPU and 20 GB of RAM.

\begin{table}[tb]
  \centering
  \caption{Results for detecting misuse of cryptographic libraries.}
  \label{tab:ssl_res}
  \begin{tabular}{|l|c|c|c|c|}\hline
                  & \textbf{Analysis time [ms]} & \textbf{\# Nodes} & \textbf{\# Functions} & \textbf{Problem found} \\\hline
    \multicolumn{5}{|c|}{\cellcolor{gray}Source Code}                                                                \\\hline
    Original file & 171                         & 328               & 38                    & Yes                    \\\hline
    \multicolumn{5}{|c|}{\cellcolor{gray}macOS M1 using XCode}                                                       \\\hline
    Compiled ll   & 1091                        & 5279              & 151                   & Yes                    \\\hline
    Lifted ll     & 256                         & 1743              & 76                    & Yes                    \\\hline
    Decompiled    & 179                         & 971               & 149                   & No                     \\\hline
    \multicolumn{5}{|c|}{\cellcolor{gray}Ubuntu x86-64 clang}                                                        \\\hline
    Compiled ll   & 163                         & 1371              & 57                    & Yes                    \\\hline
    Lifted ll     & 127                         & 911               & 48                    & Yes                    \\\hline
    Decompiled    & 80                          & 594               & 101                   & Yes                    \\\hline
    \multicolumn{5}{|c|}{\cellcolor{gray}Ubuntu x86-64 g++}                                                          \\\hline
    Lifted ll     & 242                         & 1702              & 89                    & Yes                    \\\hline
    Decompiled    & 148                         & 1137              & 200                   & Yes                    \\\hline
    \multicolumn{5}{|c|}{\cellcolor{gray}Linux AArch64 (cross compiled)}                                             \\\hline
    Lifted ll     & 250                         & 1891              & 93                    & Yes                    \\\hline
    Decompiled    & 158                         & 1176              & 209                   & Yes                    \\\hline
    \multicolumn{5}{|c|}{\cellcolor{gray}Linux arm 32 bit (cross compiled)}                                          \\\hline
    Lifted ll     & 132                         & 1123              & 51                    & Yes                    \\\hline
    Decompiled    & 71                          & 626               & 102                   & Yes                    \\\hline
  \end{tabular}
\end{table}

\subsection{Case Study: Cryptographic Misuse}
\label{sec:crypto}
This case study is driven by the anticipated usages of the CPG on LLVM-IR.
First, it should enable a security analysis of the LLVM-IR without the need to
rewrite existing analyses. Second, it should be scalable by introducing a
minimal number of nodes.
%and, contrary to priori work, introduce a minimal number of nodes to ensure the
%scalability.
The toolchain should be able to operate on LLVM-IR emitted during
the compilation of a program (subsequently, we call this ``compiled LLVM-IR'')
or when lifting a binary (we call this ``lifted LLVM-IR'').
To show that these properties are fulfilled, we
1) compare the sizes of graphs retrieved from compilers and lifters,
2) compare the runtime of the analysis,
and 3) show that the weakness can be identified with the same analysis in
all samples.

We implemented a TLS-client in C++ which uses the \texttt{openssl} library. It
accepts the insecure hashing algorithm MD5 as one of the options.
First, we tested the toolchain against the original cpp file, which
identified the respective issue. Next, we used XCode on macOS with the M1 chip
and clang on Ubuntu running on a x64 CPU to emit the LLVM-IR which can be
retrieved during compilation.
As LLVM-IR also serves as target LLVM-IR for many lifters, we lifted binaries
of the test file which had been compiled on the Mac and on Ubuntu with various
compilers.
We use RetDec \cite{RetDec} to lift the binaries to LLVM-IR and also
decompiled them to a C-style file\footnote{We compiled a custom version
  of RetDec to update the disassembler and support the \texttt{endbr64}
  instruction which had not been supported at the time of the experiments.}.
%To assess the applicability of the toolchain for multiple compilers and
%architectures, we compiled the binaries with different compilers and for
%different architectures: on macOS with the M1 chip (AArch64 architecture), we
%used XCode to compile the binary and emit LLVM-IR, on Ubuntu running on a x86-64
%CPU, we used clang and g++.
Table \ref{tab:ssl_res} summarizes
%summarizes some statistics on the graph for the respective tests. In particular,
%it shows how long the analysis took,
the analysis time, how many nodes and functions are included
in the graph and if the problem could be found successfully. We discuss the
observations %in-depth
in the following paragraphs.

%Unfortunately, the last stable version of RetDec failed to lift and decompile
%some of the files correctly. We therefore compiled the latest commits from
%github to lift and decompile the binary. However, we found that for the binaries
%compiled on x86 using gcc, the \texttt{endbr64} instruction had not been handled
%by RetDec which required an upgrade of Capstone together with minor adaptions of
%RetDec\footnote{We will open a pull request to RetDec to include the required
%changes.}.

\noindent\textbf{Size of the graphs.}~
One of our goals is to keep the sizes of the graph small. Therefore, we compare
the size of the graphs retrieved from compiled and lifted LLVM-IR
and when decompiling a binary file.

One observation is the significant increase in functions contained in the
LLVM-IR compared to the original C file.
% for both, the compiled and lifted LLVM-IR.
This can be explained by stubs introduced
by the compiler. Note, however, %However, it is notable to mention
that RetDec seems to remove
some of the functions which have been introduced during compilation. This
reduction facilitates and speeds up a subsequent security analysis on the
resulting graph.

Not only does RetDec reduce the number of functions contained
in the binary but it also reduces the number of nodes compared to compiled
LLVM-IR. This observation is in-line with recent research which found that some
lifters, including RetDec, can reduce the complexity of the code
represented by LLVM-IR as well as the number of elements it contains
\cite{liu2022sp} while keeping the main functionality of the code available.
The authors further observed that RetDec's output is not suitable for
recompiling in most cases.
%The authors further argue that the LLVM-IR retrieved by
%lifting using RetDec can support a security analysis but is not suitable for
%recompiling binaries in most cases.
However, as the CPG
library aims to handle incomplete, non-compilable and to a certain extent even
incorrect code, this limitation should not affect the representation and further
analysis.

Compared to the lifted LLVM-IR, the decompiled C files %produced by RetDec
contain more functions but less nodes. This is
explained by the possibility to summarize multiple LLVM-IR instructions in a
single C statement.
Overall, the reduction of nodes can be explained by RetDec's passes which aim to
eliminate unnecessary code.

\noindent\textbf{Runtime of the analysis.}~
We ran the translation to the CPG and the bug detection query 100 times for each
of the files and report the average runtimes in Table \ref{tab:ssl_res}.
First, it is interesting to note that the analysis time of the decompiled files
is comparable to the one of the original cpp-file. The reduced number of nodes
explains the speedup in some cases.
The overall analysis time for the LLVM-IR files is
%comparable to than the one of the original cpp-file
ranging between 0.74
to 11.1 times the time of the original file. It is notable that the graphs
retrieved from the LLVM-IR files contain 2.8 to 16.1 times the amount of nodes
of the original file and still the runtime improved.

\noindent\textbf{Identification of weaknesses.}~ To detect the misconfiguration
in the test file, we implemented a query to identify the arguments of calls to
the function \texttt{SSL\_CTX\_set\_cipher\_list}. To implement this analysis,
we use the constant propagation implemented in the analysis module included in
the CPG
library\footnote{\url{https://github.com/Fraunhofer-AISEC/cpg/tree/master/cpg-analysis}}.

With the query, we are able to identify the flaw in the original C file and in
the compiled and lifted LLVM-IR files.
However, when
decompiling the binary compiled on macOS using the M1 chip,
we failed to identify the misuse. We manually investigated the case and found
that the CDT library\footnote{\url{https://www.eclipse.org/cdt/}} which the CPG
library uses for parsing the C file fails to identify the name of a field
correctly. Therefore, the data flow between the field and the method call is not
resolved.

\noindent\textbf{Stability of the translation.}~
All samples could be represented in the CPG without crashes. However, the
LLVM-IR retrieved during compilation of a program contains a much richer
semantics and uses various different instructions.
This results in warnings, some of which show that nested instructions
are not yet handled. The other ones indicate that a different scoping for
variables in a try-catch block is expected because LLVM-IR's scoping differs
to other languages.
%This results in several
%warnings or errors reported during the translation as some properties expected by the CPG
%library do not hold in LLVM-IR. The LLVM-IR retrieved on macOS
%results in a total of 53 warnings, whereas the one retrieved on Ubuntu results
%in 14 warnings. The vast majority of the warnings originate from handling
%try-catch statements. The warnings indicate that one of the following CPG passes
%expects a different scope of validity for a variable. In the case of a try-catch
%statement, a variable declared inside the try-block would not be available
%outside this block. However, LLVM-IR does not encode such a scoping which results in a warning of the CPG pass.

\begin{table}[tb]
  \centering
  \caption{Performance when analyzing Rust libraries.}
  \label{tab:rust}
  \resizebox{\textwidth}{!}{
    \begin{tabular}{|l|l|c|c|c|c|c|}\hline
      \# & \textbf{Filename}            & \textbf{LoC} & \textbf{\# Nodes} & \textbf{\# Functions} & \textbf{\# Errors} & \textbf{Analysis time [ms]} \\\hline
      1  & addr2line                    & 879          & 2327              & 29                    & 9                  & 3641                        \\\hline
      2  & adler                        & 488          & 1707              & 25                    & 2                  & 507                         \\\hline
      3  & alloc                        & 4925         & 13482             & 253                   & 91                 & 6505                        \\\hline
      4  & cfg\_if                      & 9            & 1                 & 0                     & 0                  & 23                          \\\hline
      5  & compiler\_builtins           & 9990         & 34304             & 338                   & 0                  & 23670                       \\\hline
      6  & core                         & 80193        & 263729            & 3608                  & 1879               & 2872096                     \\\hline
      7  & gimli                        & 23702        & 72845             & 411                   & 43                 & 112269                      \\\hline
      8  & hashbrown                    & 276          & 529               & 26                    & 0                  & 193                         \\\hline
      9  & libc                         & 1477         & 3619              & 130                   & 0                  & 646                         \\\hline
      10 & memchr                       & 11063        & 40602             & 257                   & 108                & 32639                       \\\hline
      11 & miniz\_oxide                 & 15760        & 54868             & 294                   & 166                & 79863                       \\\hline
      12 & object                       & 14174        & 50060             & 277                   & 5                  & 47806                       \\\hline
      13 & panic\_abort                 & 71           & 87                & 9                     & 0                  & 124                         \\\hline
      14 & panic\_unwind                & 927          & 2619              & 67                    & 25                 & 610                         \\\hline
      15 & proc\_macro                  & 92115        & 244010            & 5488                  & 2570               & 15260350                    \\\hline
      16 & rustc\_demangle              & 14669        & 44069             & 437                   & 309                & 43281                       \\\hline
      17 & rustc\_std\_workspace\_alloc & 9            & 1                 & 0                     & 0                  & 107                         \\\hline
      18 & rustc\_std\_workspace\_core  & 9            & 1                 & 0                     & 0                  & 102                         \\\hline
      19 & std                          & 157377       & 468223            & 5923                  & 2629               & 9303378                     \\\hline
      20 & std\_detect                  & 558          & 1921              & 15                    & 0                  & 659                         \\\hline
      21 & unwind                       & 106          & 230               & 2                     & 0                  & 273                         \\\hline
    \end{tabular}}
\end{table}

\subsection{Application to the Rust Runtime}
To assess the applicability to real-world programs, we retrieved the LLVM-IR
from the standard and core libraries of Rust. We chose Rust since it is
%one of the programming languages which is
not yet supported by the CPG implementation and provides the option to compile
to LLVM-IR. Overall, the test set includes 21 distinct LLVM files which are
listed in Table \ref{tab:rust} together with their size and the results. We
report the time it took to translate the file (including various CPG passes) as
well as the number of nodes which could not be parsed accurately. For the
latter, we need to extend the LLVM-specific translation to include more cases of
``nested'' LLVM expressions.

\begin{figure}[tb]
  \centering
  \resizebox{0.9\textwidth}{!}{
    \begin{tikzpicture}
      \begin{axis}[
        xlabel={Nodes/LoC},
        ylabel={ProblemNodes/Nodes [\%]},
        %ymode=log,
        %xmode=log
        ]

        \addplot[color=blue, mark=*, only marks]
        coordinates {(2.647326507,0.3867641) (3.49795082,0.1171646)
            (2.7374619,0.674974) (0.11111111,0) (3.433833934,0) (3.288678563,0.71247379)
            (3.073369336,0.05902944) (1.91166667,0) (2.450236967,0)
            (3.670071409,0.265996749) (3.481472081,0.30254428811)
            (3.531818823,0.00998801438) (1.225352113, 0) (2.825242718,0.95456281023)
            (2.648971394,1.05323552313) (3.0042266,0.70117316027) (0.11111111,0)
            (0.11111111,0) (2.97516791,0.56148459174) (3.44265233,0) (2.169811321,0)};
      \end{axis}

      \begin{axis}[
        xlabel={LoC},
        ylabel={ProblemNodes/Nodes [\%]},
        xticklabel pos=right,
        xlabel near ticks,
        legend pos=outer north east
        ]
        % A bit hacky but fits the upper plot. Only here to get the legend...
        \addplot[color=blue, mark=*, only marks] coordinates {(1,0)};
        \addlegendentry{Nodes/LoC vs. ProblemNodes/Nodes};

        \addplot[color=red, mark=o, only marks]
        coordinates {(879,0.3867641) (488,0.1171646)
            (4925,0.674974) (9,0) (9990,0) (80193,0.71247379)
            (23702,0.05902944) (276,0) (1477,0)
            (11063,0.265996749) (15760,0.30254428811)
            (14174,0.00998801438) (71, 0) (927,0.95456281023)
            (92115,1.05323552313) (14669,0.70117316027) (9,0)
            (9,0) (157377,0.56148459174) (558,0) (106,0)};
        \addlegendentry{LoC vs. ProblemNodes/Nodes};

      \end{axis}

    \end{tikzpicture}}
  \caption{Relation between lines of code, nodes in the CPG and the fraction
    of ProblemNodes. For non-trivial samples, the error-rates are randomly
    distributed.}
  \label{fig:errors}
\end{figure}
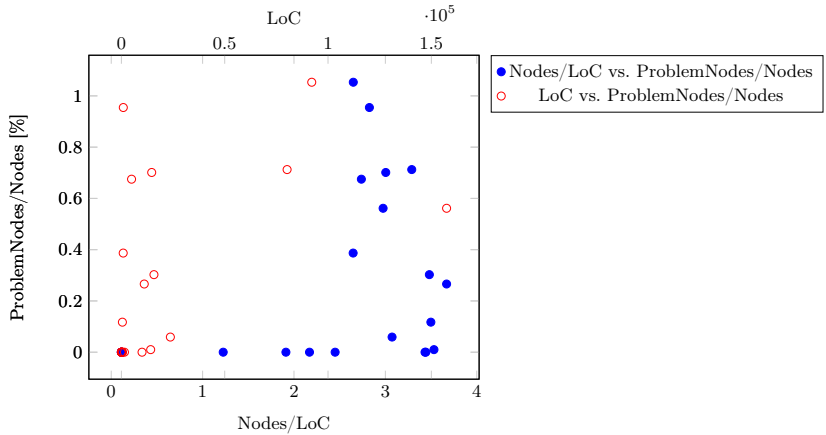

\noindent\textbf{Stability.}~
We want to assess the maturity level of the translation step
against a large and unknown codebase consisting of a total of $428,777$ lines of
LLVM-IR. To measure this, the graph includes specific nodes, called
\texttt{ProblemNode}, for each expression which could not be parsed correctly.
While we handle all types of instructions, some arguments of the instructions
can be computed in line by type casts, or simple arithmetic operations, among
others. Overall, we could observe $7,836$ of such \texttt{ProblemNode}s, which
accounts for $0.60\%$ of all $1,299,234$ nodes. This result is encouraging and
indicates that the current implementation is already capable of handling the vast
majority of all combinations of statements\footnote{We will manually
  investigate the ProblemNodes to parse the statements in the future.}.

The fraction of nodes which cannot be handled differs significantly among the
samples and ranges between $0\%$ to $1.05\%$. Larger files are more likely to
lead to an error during the translation. In addition, it is possible that the
varying amount of complexity of the code could trigger more errors. To validate
this, we set the average number of CPG nodes per line of code as complexity of
the LLVM instructions. Among the samples, this ratio ranges between $1.22$ and
$3.67$.

We plot this relation in Figure \ref{fig:errors}. Neither of the graphs
gives a strong indication for this idea since the error rates seem to be
randomly distributed for all non-trivial samples. Neither the size nor the
complexity of the samples lead to a conceptual limitation. Instead, some samples
use unsoppurted expressions more frequently which can easily be addressed in the
implementation.

\begin{figure}[tb]
  \centering
  \resizebox{0.6\textwidth}{!}{
    \begin{tikzpicture}
      \begin{axis}[
          xlabel={\# Nodes in the CPG},
          ylabel={Analysis time [s]},
          ymode=log,
          %xmode=log,
          legend pos=south east
        ]

        \addplot[color=blue, mark=square]
        coordinates {(1,0.023) (1,0.102) (1,0.107) (87,0.124) (230,0.273)
            (529,0.193) (1707,0.507) (1921,0.659) (2327,3.641) (2619,0.610) (3619,0.646)
            (13482,6.505) (34304,23.670) (40602,32.639) (44069,43.281) (50060,47.806)
            (54868,79.863) (72845,112.269) (244010,15260.350) (263729,2872.096)
            (468223,9303.378)};
        \addlegendentry{Including CPG passes};

        \addplot[color=red, mark=o]
        coordinates {(1,0.012) (1,0.181) (1,0.519) (87,0.137) (230,0.22)
            (529,0.129) (1707,0.885) (1921,0.349) (2327,5.108) (2619,0.296) (3619,0.404)
            (13482,2.057) (34304,5.203) (40602,3.828) (44069,7.850) (50060,3.846)
            (54868,10.354) (72845,24.883) (244010,248.362) (263729,132.879)
            (468223,515.532)};
        \addlegendentry{Only LLVM-related CPG pass};
      \end{axis}
    \end{tikzpicture}}
  \caption{Analysis time vs. \# Nodes. Note the logarithmic y scale.}
  \label{fig:at_nodes}
\end{figure}
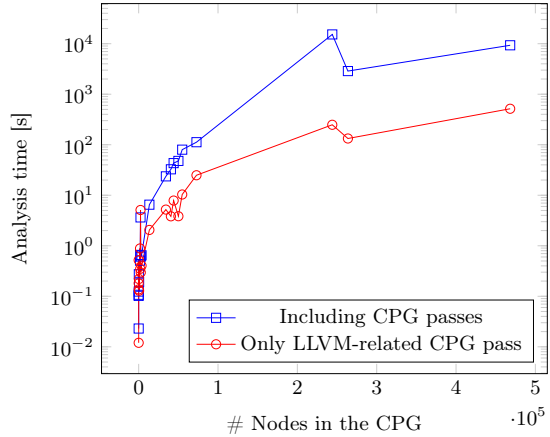

\noindent\textbf{Scalability.}~
Another goal is to assess the scalability of the implementation on
real-world software with many lines of code. Two factors can impact the analysis
time: The lines of code and the number of nodes in the graph. According to Table
\ref{tab:rust}, an increase of LoC leads to more nodes in the graph in most
cases. Figure~\ref{fig:at_nodes} plots the time of the analysis (i.e., the
translation to the CPG and all CPG passes but the
\texttt{ControlFlowSensitiveDFGPass}) for the number of nodes. With the
exception of one sample, the analysis time seems to grow linearly depending on
the number of nodes in the graph. Interestingly, when we only consider the
analysis time of the LLVM-specific translation and pass of the CPG, the outlier
is no longer present. This shows that the LLVM-related translation and CPG pass
do scale well even for larger samples but that some of the other CPG passes seem
to perform poorly in the presence of a specific combination of nodes.

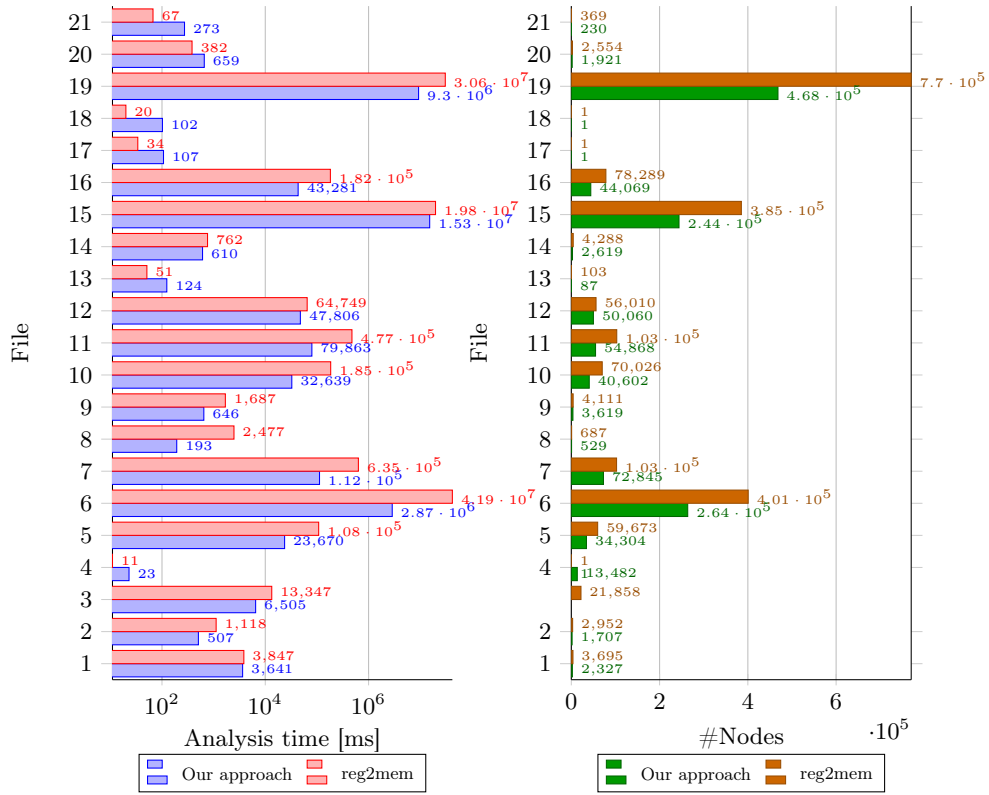
\begin{figure}[h!]
  \centering
  \begin{minipage}{.5\textwidth}
    \centering
    \begin{tikzpicture}
      \begin{axis}[
          xbar=0pt,
          width  = \textwidth,
          axis y line*=left,
          axis x line=bottom,
          height = 300pt,
          %major x tick style = transparent,
          enlarge y limits=0.025,
          bar width=5pt,
          xmajorgrids = true,
          ylabel = {File},
          xlabel = {Analysis time [ms]},
          %symbolic y coords={1,2,3,4,5, 6, 7, 8, 9, 10, 11, 12, 13, 14, 15, 16,
          %    17, 18, 19, 20, 21},
          ytick = data,
          xmode=log,
          scaled y ticks = true,
          axis line style={-},
          legend columns=2,
          legend cell align=left,
          legend style={
              at={(0.5,-0.11)},
              anchor=north,
              column sep=1ex,
              nodes={scale=0.75, transform shape}
            },
          nodes near coords,
          nodes near coords style={font=\tiny},
          nodes near coords align=horizontal,
          point meta=rawx
        ]
        \addplot+[xbar]
        coordinates {(3641, 1) (507, 2) (6505, 3) (0023, 4) (23670, 5) (2872096, 6)
            (112269, 7) (0193, 8) (0646, 9) (32639, 10) (79863, 11) (47806, 12) (0124, 13)
            (0610, 14) (15260350, 15) (43281, 16) (0107, 17) (0102, 18)
            (9303378, 19) (0659, 20) (0273, 21)};

        \addplot+[xbar]
        coordinates {(3847, 1) (1118, 2) (13347, 3) (0011, 4) (108124, 5)
            (41934277, 6) (635146, 7) (2477, 8) (1687, 9) (184776, 10) (476777, 11)
            (64749, 12) (0051, 13) (0762, 14) (19795079, 15) (182274, 16) (0034, 17)
            (0020, 18) (30598095, 19) (0382, 20) (0067, 21)};
        \legend{Our approach, reg2mem}
      \end{axis}
    \end{tikzpicture}
  \end{minipage}%
  \begin{minipage}{.5\textwidth}
    \begin{tikzpicture}

      \begin{axis}[
          xbar=0pt,
          width=\textwidth,
          %axis y line*=right,
          axis x line=bottom,
          height=300pt,
          %major x tick style = transparent,
          %ylabel near ticks,
          bar width=5pt,
          enlarge y limits=0.025,
          xmajorgrids=true,
          xlabel={\#Nodes},
          ylabel={File},
          %symbolic x coords={1,2,3,4,5, 6, 7, 8, 9, 10, 11, 12, 13, 14, 15, 16,
          %    17, 18, 19, 20, 21},
          ytick=data,
          axis line style={-},
          nodes near coords,
          node near coords style={font=\tiny},
          nodes near coords align=horizontal,
          point meta=rawx,
          legend columns=2,
          legend cell align=left,
          legend style={
              at={(0.5,-0.11)},
              anchor=north,
              column sep=1ex,
              nodes={scale=0.75, transform shape}
            },
        ]

        \addplot+[xbar, style=black!60!green, fill=black!40!green]
        coordinates {(2327, 1) (1707, 2) (13482, 4) (1, 4) (34304, 5) (263729, 6)
            (72845, 7) (529, 8) (3619, 9) (40602, 10) (54868, 11) (50060, 12)
            (87, 13) (2619, 14) (244010, 15) (44069, 16) (1, 17) (1, 18)
            (468223, 19) (1921, 20) (230, 21)};

        \addplot+[xbar, style=black!40!orange, fill=black!20!orange]
        coordinates {(3695, 1) (2952, 2) (21858, 3) (1, 4) (59673, 5) (400775, 6)
            (102504, 7) (687, 8) (4111, 9) (70026, 10) (102923, 11) (56010, 12)
            (103, 13) (4288, 14) (385270, 15) (78289, 16) (1, 17) (1, 18)
            (770152, 19) (2554, 20) (369, 21)};
        \legend{Our approach, reg2mem}
      \end{axis}

      %    \begin{axis}[
      %        width  = \textwidth,
      %        axis y line*=left,
      %        axis x line=bottom,
      %        height = 9cm,
      %        major x tick style = transparent,
      %        enlarge x limits=0.02,
      %        bar width=5pt,
      %        ymajorgrids = true,
      %        ylabel = {Analysis time [s]},
      %        symbolic x coords={1,2,3,4,5, 6, 7, 8, 9, 10, 11, 12, 13, 14, 15, 16, 17, 18, 19, 20, 21},
      %        xtick = data,
      %        ymode=log,
      %        scaled y ticks = true,
      %        axis line style={-},
      %        legend columns=2,
      %        legend cell align=left,
      %        legend style={
      %            at={(0.5,-0.15)},
      %            anchor=north,
      %            column sep=1ex
      %          },
      %      ]
      %      \addplot+[mark=*, only marks, style=blue]
      %      coordinates {(1, 3.641) (2, 0.57) (3, 6.505) (4, 0.023) (5, 23.670) (6, 2872.096)
      %          (7, 112.269) (8, 0.193) (9, 0.646) (10, 32.639) (11, 79.863) (12, 47.806) (13, 0.124)
      %          (14, 0.610) (15, 15260.350) (16, 43.281) (17, 0.107) (18, 0.102) (19,
      %          9303.378) (20, 0.659) (21, 0.273)};
      %
      %      \addplot+[mark=square*, only marks, style=red]
      %      coordinates {(1, 3.847) (2, 1.118) (3, 13.347) (4, 0.011) (5, 108.124) (6,41934.277)
      %          (7, 635.146) (8, 2.477) (9, 1.687) (10, 184.776) (11, 476.777) (12, 64.749) (13, 0.051)
      %          (14, 0.762) (15, 19795.079) (16, 182.274) (17, 0.034) (18, 0.02) (19,
      %          30598.095) (20, 0.382) (21, 0.067)};
      %      \legend{Time with our approach, Time with reg2mem}
      %    \end{axis}

    \end{tikzpicture}
  \end{minipage}
  \caption{Performance comparison of our approach and prior work. The analysis time
    and the number of nodes are typically much smaller with our improvements.}
  \label{fig:comparison}
\end{figure}

\noindent\textbf{Comparison to prior work.}~
To compare our approach to prior work which relied on LLVM's reg2mem pass to
remove $\varphi$ nodes, we ran our toolchain but first executed the respective
pass.
As Figure \ref{fig:comparison} shows, our approach leads to a significant
reduction of nodes and time required to generate the graph.

\section{Discussion}
\label{sec:discussion}

Our evaluation suggests that our translation and CPG model can
unify source code and low-level representations such as
LLVM-IR in a single graph representation. This increases the reusability of
analyses and queries on the graph.

We found that the LLVM-IR retrieved from binary lifters is significantly easier
to handle by the graph. This is due to the fact that most lifters tend to use
rather conservative steps for their translation. This results in the LLVM-IR being
closer to assembly code with comparably simple types of instructions. The LLVM-IR
retrieved during the compilation, in contrast, features numerous
highly specialized instructions which typically make the translation more
difficult. Furthermore, the graphs retrieved from lifted binaries
are typically smaller than the ones which can be retrieved when the LLVM-IR is
retrieved during the compilation. This makes it an interesting application since
it simplifies and speeds up the analysis. Last, we found that the graph of the
decompiled binary is only marginally smaller than the one holding the lifted
LLVM-IR instructions. This small advantage will, however, not outweigh the
error-prone and time-consuming decompliation step in most scenarios which is
required to retrieve the code.

%\subsection{Threats to Validity}
\noindent\textbf{Validity of the Results.}~
The main threat to the validity of the findings is the set of test samples. In
particular, as we could see in Section \ref{sec:crypto}, the compiler has a
significant impact on the generated LLVM-IR and the resulting complexity which needs
to be handled by our toolchain. Hence, testing the toolchain against different
compilers and configurations might lead to different results. To address this
potential issue, we used XCode on macOS and clang on Ubuntu,
and we also generated the LLVM-IR with Rust's crates build system.
Furthermore, we used a binary lifter to showcase a possible application to such a
scenario.

%\subsection{Limitations}
\noindent\textbf{Limitations.}~
As our evaluation against the Rust standard library showed, a small amount of
instructions could not be parsed correctly. This is explained by the possibility
of LLVM-IR to hold sub-statements for the arguments. While we do handle the
concepts and operators (e.g., casts), their
potential usage in a specific sub-statement needs to be added to
the translation step. To identify all possible combinations, a more extensive
testing is required.

%\subsection{Future Work and Research Directions}
\noindent\textbf{Future Work and Research Directions.}~
The resulting graph can be used as an entry point for further research to better
include specifics of certain platforms.
One example is the analysis of the LLVM-IR emitted by XCode for apps written in
Apple's programming languages Swift or Objective-C. Their calling
conventions differ significantly from other programming languages. As an
example, Objective-C makes use of a dynamic dispatching routine which requires
extensive tracing of a method's arguments to recover type information and the
method name as a string \cite{schuette2019lios,egele2011pios}. This
information is present in the CPG but has to be combined to identify
the calls. Similarly, it is necessary to model Swift's calling conventions and
memory model since it differs significantly from the one of C++
\cite{tiganov2020swan,kraus2018the}. However, to date, the differences are
not fully explored. Future work should identify differences and
integrate this knowledge into the CPG. %The general lack of analysis tools for
%these platforms require extensive research in the area.

Furthermore, software written in C or C++ can rely on macros which are used
similar to function calls in the source code and represented as such but are
replaced with their specific implementation in LLVM-IR. This discrepancy needs
to be addressed appropriately to better analyze such programs. In the current
stage, addressing such inconsistencies between source code and the binary is
left to manual efforts of the user of the cpg library. Additional efforts
are necessary to reduce these manual efforts and ease the usability of the
analysis toolchain.

%Last, it is promising to adapt the solution to challenges arising from different
%scenarios analyzing closed-source software such as firmware of IoT devices.
Last, adapting the solution to the analysis of closed-source software is
promising.
%However, this requires an extensive evaluation of different binary
%lifters and their applicability to real-world software.
Recent research \cite{liu2022sp} showed that lifting is a stable technique for many
applications. However, lifted or decompiled binaries still
suffer from a lack of information which are crucial for a security analysis
\cite{mantovani2022the}. Hence, further research should %aim to recover even
%more information from the binary files and
study which gaps still exist to apply existing tools to lifted binaries.

%\subsection{Generalizability}
\noindent\textbf{Generalizability.}~
Since the SSA form is also used by other IRs (e.g. Shimple \cite{shimple}, WALA
\cite{wala}, SIL \cite{sil}), some of the challenges generalize to those
IRs. %Therefore, in case a user wants to add further IRs or programming languages
%using the SSA form to the CPG, she can reuse the concepts presented in this
%paper.
Hence, the concepts presented in this paper can be reused to add further code
representations using the SSA form to the CPG.
Furthermore, some parts of our concept could be ported to other projects
which suffer from similar issues. However, the applicability and impact depend
on the projects' data models.
%As the efforts of implementing the specific frontends are highly specific
%to the IR and the tool in which they are integrated, the generalizability to
%other IRs and platforms is limited.

\section{Conclusion}
\label{sec:conclusion}
We showed how we extended an open source CPG implementation to handle
LLVM-IR. While the majority of instructions can easily be mapped to the high-level
equivalents, the $\varphi$ instruction and the LLVM exception handling
instructions impose challenges to the translation. However, we could transform
the program to the CPG representation with a reasonable increase in nodes
while prior work suffered from huge performance penalties.
The similarity between the resulting graph and the one of the code fractions in
high-level languages allows to reuse existing analyses detecting security
weaknesses or bugs. Our evaluation suggests that the approach scales to larger
%the resulting graph is suitable for a subsequent security analysis which can
%reuse wide parts of the graph queries used for source code and scales to larger
projects. Future work is necessary to include characteristics of some
programming languages (e.g. Swift), to add analyses for further use cases,
%which had previously not been promising on the CPG,
%and to better assess which steps are necessary to apply the toolchain to binary lifting.
and to study the gaps of binary lifting.

\textbf{Acknowledgements.}~
This work was partially funded by the Horizon 2020 project MEDINA, grant agreement ID 952633.

%\balance
%\bibliographystyle{IEEEtranS}
\bibliographystyle{splncs04}
\urlstyle{tt}
\bibliography{biblio}

\end{document}